\documentstyle[11pt,aasms4]{article}

\begin{document}

\title{Kinematics of Molecular Hydrogen Emission from Pre-planetary
Nebulae: RAFGL 2688 and RAFGL 618} 
\author{Joel H. Kastner\altaffilmark{1}, David
A. Weintraub\altaffilmark{2}, Ian Gatley\altaffilmark{1}, and LeeAnn
Henn\altaffilmark{3}} 

\altaffiltext{1}{Chester F. Carlson Center for Imaging Science, Rochester
Institute of Technology, 54 Lomb Memorial Dr., Rochester, NY
14623-5604 USA; jhk@cis.rit.edu}

\altaffiltext{2}{Department of Physics \& Astronomy,
Vanderbilt University, P.O. Box 1807 Station B, Nashville, TN 37235}

\altaffiltext{3}{Center for Space Research, Massachusetts Institute of
Technology, NE80-6007, Cambridge, MA 02139 USA}

\begin{abstract}

We present high spectral resolution maps of near-infrared molecular hydrogen
emission from the bipolar pre-planetary nebulae RAFGL 2688 and RAFGL 618,
obtained with the NOAO Phoenix spectrometer. The measured velocity
gradients along the polar axes of both nebulae indicate 
that the highest velocity gas lies closest to the central stars. 
These results support the suggestion
that the polar H$_2$ emission regions of both nebulae contain shocked gas
formed as fast ($\sim50-150$ km s$^{-1}$), collimated, post-asymptotic giant
branch (AGB) winds collide with slower-moving ($\sim10-20$ km s$^{-1}$)
material previously ejected while
the central stars were still on the AGB. The kinematics of H$_2$ emission
perpendicular to the polar axis of RAFGL 2688 are consistent with
a model combining expansion along the equator at $5-10$ km s$^{-1}$ with
rotation about the polar axis at $5-10$ km s$^{-1}$. 
The rapid onset of the common envelope phase of a
close binary system may explain both the bipolar structure of RAFGL 2688 and
the presence and complex kinematics of its shocked H$_2$ emission.

\end{abstract}

\keywords{pre-planetary nebulae: individual (RAFGL 2688, RAFGL 618) ---
pre-planetary nebulae: molecules --- stars: mass loss --- stars: AGB and
post-AGB --- ISM: dust}

\section{Introduction}

The bulk of the mass lost by stars of initial mass $\sim1-8$ $M_\odot$ is
shed while such stars are on the asymptotic giant branch (AGB).  Evidence is
fast accumulating that, during late AGB or post-AGB evolutionary stages, the
mass loss geometry of such stars changes from more or less spherically
symmetric to axially symmetric, with the result that the vast majority of
planetary nebulae (PNs) --- the endpoints of post-main sequence stellar
evolution for intermediate-mass stars --- exhibit axisymmetric structures,
ranging from elliptical to bipolar (see, e.g., reviews compiled
by Kastner, Soker, \& Rappaport 2000). Planetaries in the latter structural
category are very likely to possess molecular envelopes (presumably the
remnants of the ejected AGB envelope) that are readily detectable in the
near-infrared rovibrational lines of H$_2$ (Kastner et al.\ 1996). The
available data suggest, furthermore, that the onset of near-infrared H$_2$
emission in PNs can be traced back to the pre-planetary nebula (PPN) phase
but {\it not} back to the AGB phase of evolution (Weintraub et al.\
1998). These observations suggest that further studies of H$_2$ emission
from PPNs and PNs may offer insight into the transition from AGB star to PN
and from spherical to axisymmetric mass loss.

The Egg Nebula (RAFGL 2688; Ney et al.\ 1975) is perhaps the best-known
example of such a transition object. Recent {\it Hubble Space Telescope}
(HST) optical and near-infrared images of the Egg reveal a striking
juxtaposition of circularly and axially symmetric structures (Sahai et al.\
1998a, b). This puzzling combination of structures has
been detected in HST 
images of several other PPNs and PNs (see, e.g., reviews
compiled by Kastner, Soker, \& Rappaport 2000). An outstanding
feature of RAFGL 2688, however, is the ``quadrapolar'' morphology of 2.122
$\mu$m H$_2$ emission detected in near-infrared images (Gatley et al.\
1988); this structure is seen most clearly in an HST/NICMOS image (Fig.\
\ref{fig:2688NICMOS}; Sahai et al.\ 1998a). This image shows that the
linear extents and bow-shock-like structures of the polar and equatorial
plane H$_2$ emission regions are very similar, suggesting a common
origin. Furthermore, the sharp, limb-brightened boundaries of the H$_2$
emission in all four ``lobes'' is suggestive of a fairly sudden event in the
object's recent past.

The kinematics of molecular emission from RAFGL 2688 have been the subject of
numerous investigations.  Bieging \& Nguyen-Q-Rieu (1988,
1996) and Cox et al.\ (2000) mapped the 
velocity distribution of rotational emission of various species with radio
interferometers, and Cox et al.\ (1997) imaged near-IR rovibrational H$_2$
emission at moderate velocity resolution.  Radio interferometric
observations of HCN by Bieging \& Nguyen-Q-Rieu yielded the rather
surprising result of velocity gradients both along {\it and perpendicular
to} the polar axis of the system, a result confirmed by Cox
et al. These gradients,
which are of similar magnitude, were interpreted by Bieging \& Nguyen-Q-Rieu
(1988, 1996) as arising out of a combination of spherical expansion and
rotation about the polar axis. Cox et al.\ (1997, 2000) and
Lucas et al.\ (2000), however, intepret the H$_2$
kinematics of RAFGL 2688 in terms of a system of multiple jets.

The evolved bipolar nebula RAFGL 618 (Westbrook et al.\ 1975) might be
considered the direct evolutionary descendent of RAFGL 2688 in many
respects. Both objects display molecule-rich circumstellar envelopes,
carbon-rich circumstellar chemistries, and dusty, bipolar reflection
nebulosities. Whereas RAFGL 2688 harbors an F-type central star and is not
known to contain an H II region, RAFGL 618 possesses a B-type central star
and compact H II region, suggesting that the latter nebula has evolved
farther toward the PN stage. Like the Egg Nebula, RAFGL 618 displays bright
near-infrared H$_2$ emission, but in the case of RAFGL 618 the H$_2$ emission
appears to be confined to a region along its polar axis (Latter et al.\
1995). 

While the ``quadrapolar'' structure of the H$_2$ emission in the Egg appears
to be unique, it may be a short-lived evolutionary phase common to many or
most AGB stars as they go through the transition to PPNs.  In RAFGL 618, 
however, the H$_2$ morphology also is unusual as emission is
found only along the polar axes, whereas most PPN in which
H$_2$ is detected exhibit the strongest H$_2$ along the
equatorial plane (Kastner et al. 1996). Thus, the kinematics and  
morphology of H$_2$ emission from these two objects are strongly
deserving of more careful study. To investigate the H$_2$
velocity gradients of RAFGL 2688 in detail, we 
obtained a series of high-resolution, long-slit spectra in the 2.12 $\mu$m
region with the NOAO\footnote{National Optical Astronomy Observatories is
operated by Associated Universities for Research in Astronomy, Inc., for the
National Science Foundation.} Phoenix near-infrared spectrometer (Hinkle et
al.\ 1998). We also obtained a Phoenix spectrum of 2.12 $\mu$m H$_2$
emission from RAFGL 618. As we now describe, examination and comparison of
the H$_2$ kinematics of these two objects, as revealed in the Phoenix
spectroscopic data, offer clues to their detailed structures and recent
evolutionary histories.

\section{Observations}

Data presented here were obtained with Phoenix on the 2.1 m telescope at
Kitt Peak, AZ, in 1997 June (RAFGL 2688) and 1997 December (RAFGL
618). Phoenix illuminates a $256\times1024$ section of an Aladdin InSb
detector array. The spectrograph slit was $\sim60''\times1.4''$ 
oriented approximately east-west. The pixel scale along the dispersion axis
of the spectrograph was 1.29 km s$^{-1}$. From the widths of OH airglow
lines present in the raw spectra, we estimate the velocity resolution as
$\sim4$ km s$^{-1}$ at the time these spectra were obtained. This resolution
is comparable to that of the mm-wave molecular line interferometry (e.g.,
Bieging \& Nguyen-Q-Rieu 1996), and represents an order of magnitude
improvement over previous imaging spectroscopy of H$_2$ emission from RAFGL
2688 (Cox et al.\ 1997).  The pixel scale in the cross-dispersion direction
was $0.35''$, and the spatial resolution was $\sim1.5''$.

A spectral image centered near the 2.1218 $\mu$m $S(1)$, $v=1-0$
transition of H$_2$ was obtained at each of 12 spatial positions as the slit
was stepped from south to north across RAFGL 2688. The step size, $1.0''$,
provided coverage of the entire H$_2$ emitting region (Fig.\ 1) with spatial
sampling approximating the slit height. For RAFGL 618, whose bright H$_2$
emission regions are oriented almost perfectly east-west (Latter et al.\
1995) --- parallel to the Phoenix slit --- we obtained a single spectral image
centered on the object. Integration times for both sources were 1200 sec per
position. Spectral images were reduced and wavelength calibrated as
described by Weintraub et al.\ (1998). Bad pixels were removed by
calculating the running mean for $7\times7$ groups of pixels and replacing
outliers in each pixel group with the running mean. For the RAFGL 2688 data,
the reduced spectral images were stacked in declination according to the
commanded telescope offsets, to produce a (RA, dec, velocity) data cube of
H$_2$ emission.

\section{Results}

\subsection{RAFGL 2688}

\subsubsection{H$_2$ position-velocity maps and selected spectra}

Calibrated spectral images of RAFGL 2688 are presented in Fig.\
\ref{fig:2688posvel}. 
All four distinct H$_2$ emitting regions (hereafter referred to as the N, E, S,
and W lobes) seen in the NICMOS image (Fig.\ \ref{fig:2688NICMOS}) are
clearly detected in these spectral images (see, e.g., the
dec = $-$1\arcsec\ panel). In addition, the large velocity 
gradients of molecular emission in RAFGL 2688 are readily apparent in the
Phoenix data. The S lobe H$_2$ emission (top 4 panels of Fig.\
\ref{fig:2688posvel}) is predominantly redshifted with
respect to the systemic velocity of RAFGL 2688 ($V_{sys}$),
with velocities ranging 
from $\sim -5$ km s$^{-1}$ to $\sim +30$ km s$^{-1}$ relative to
$V_{sys}$. The N lobe emission (bottom 4 panels) appears almost as a
``mirror image'' of the S lobe in terms of its velocity pattern, with
emission predominantly at blueshifted velocities ranging from $\sim -35$ km
s$^{-1}$ to $\sim +5$ km s$^{-1}$ relative to $V_{sys}$. The velocity
domains of the N and S lobes are well illustrated in representative spectra
extracted from the images obtained at declination offsets of $+4''$ and
$-4''$, respectively (Fig. \ref{fig:2688spectra}a).

As expected from the known position angle (PA) of the polar axis of the
system, PA $\sim12^\circ$ (Weintraub et al.\ 2000), 
the E lobe is detected between declination
offsets of $-3''$ and $+2''$, whereas the W lobe emission is detected
between offsets of $-2''$ and $+3''$. The position-velocity images obtained
at these positions dramatically demonstrate the E-W velocity gradient
previously detected in molecular emission line studies of RAFGL 2688, as do
representative spectra extracted from the E and W lobe emission regions in
the spectral image obtained at $0''$ declination offset
(Fig. \ref{fig:2688spectra}b). That is, the E lobe is primarily blueshifted,
whereas the W lobe is primarily redshifted. As we
discuss in more detail below, however, the E-W and N-S velocity gradients 
differ in detail. In addition, the spectra
in Fig. \ref{fig:2688spectra} illustrate that the H$_2$
line profiles of the E and W lobes are generally narrower than those of the
N and S lobes.

The Phoenix spectral images also reveal velocity structure {\it within} the
N, S, and E lobes. In particular, the brightest emission from the S lobe is
seen to migrate from lower to higher redshifted velocities moving inward
(northward) from its tip at $-5''$ in dec to $-2''$ in dec.
The N lobe exhibits precisely complementary behavior, with its brightest
emission migrating from lower to higher blueshifted velocities moving inward
(southward) from $+6''$ in dec to $+2''$ in dec. The E lobe also shifts in
central velocity with declination offset, migrating from larger to smaller
blueshifted velocities moving from south to north (i.e., from $-3''$ to
$+2''$ in declination offset). In contrast, the velocity centroid of
the W lobe does not shift noticeably with declination offset.

\subsubsection{H$_2$ data cube and velocity centroids}

In Fig.\ \ref{fig:2688cube} we display selected velocity planes from the
H$_2$ data cube constructed for RAFGL 2688. The four
principal ``lobes'' of H$_2$ emission 
are apparent again, with one pair oriented parallel
(roughly N-S) and one perpendicular (roughly E-W) to the polar axis. The
velocity gradients in the N-S and E-W
H$_2$ lobe pairs are plainly evident in these images, with
the N and E lobes blueshifted by up to $\sim25$ km s$^{-1}$ and the S and W
lobes similarly redshifted. As previously mentioned, however, the N-S and
E-W H$_2$ lobe pairs differ in their detailed kinematic signatures, as we
now further describe.

The polar H$_2$ lobes of RAFGL 2688 each display velocity gradients along the
polar axis. This velocity gradient is apparent in the velocity-resolved
images of RAFGL 2688 (Fig.\ \ref{fig:2688cube}). At the most extreme
blueshifted and redshifted velocity intervals, respectively, only those
portions of the N and S lobes nearest the inferred position of the central
star are apparent, whereas the outer regions (tips) of the lobes appear
brightest at more moderate radial velocities.  This trend is made more
quantitative in a plot of the velocity centroid of H$_2$ emission as a
function of position along the polar axis (Fig.\ \ref{fig:2688poles}). To
construct this plot, we extracted spectra at each declination offset (see
caption of Fig.\ \ref{fig:2688spectra}). We then fit a Gaussian function to
the H$_2$ line profile in each spectrum, to obtain the line velocity
centroid. The resulting trend in the velocity centroids (Fig.\
\ref{fig:2688poles}) is easily anticipated based on the position-velocity
images themselves (Fig.\ \ref{fig:2688posvel}); i.e., relatively low
velocity emission ($\sim \pm 4$ km s$^{-1}$) is seen at the tips of the
lobes, while the highest velocity emission in each lobe ($\sim \pm 30$ km
s$^{-1}$) is found nearest the position of the central star. This trend is
remarkably similar to that observed in the position-velocity image of RAFGL
618 (\S 3.2).

In contrast, images extracted from the data cube (Fig.\ \ref{fig:2688cube})
demonstrate that the E lobe displays a gradient from north to south, {\it
perpendicular} to the line joining it to the central star's
position. Specifically, note that in the image centered at $-26.6$ km
s$^{-1}$, only the southernmost tip of the E lobe appears, whereas in the
image centered at $-0.7$ km s$^{-1}$, the entire E lobe is present. Hence, the
Phoenix data appear to resolve kinematically the distinct emission
components of the E lobe that are resolved spatially in the NICMOS H$_2$
image of RAFGL 2688 (Fig.\ 1).  The W lobe, meanwhile, does not reveal
similar velocity structure, a result qualitatively consistent with its
more localized appearance in the NICMOS image. 

\subsection{RAFGL 618}

The Phoenix spectral image obtained for RAFGL 618 is displayed in Fig.\
\ref{fig:618posvel}. Bright H$_2$ emission is detected along the entire
polar axis of RAFGL 618. Spectra extracted from this image (Fig.\
\ref{fig:618spectra}) demonstrate that very high velocity H$_2$ emission (up
to $\sim \pm120$ km s$^{-1}$ in each lobe) is present in this
bipolar outflow. Similarly broad line wings have been detected previously in
millimeter- and submillimeter-wave molecular spectra of RAFGL 618 (e.g.,
Gammie et al.\ 1989; Neri et al.\ 1992; Meixner et al.\
1998). The H$_2$ spectra demonstrate unequivocally that ---
as is the case for RAFGL 2688 --- 
the highest velocity molecular material is found closest to the central star
of RAFGL 618. A similar result was inferred for RAFGL 618 by
Gammie et al., based on the relative strengths of line wing emission in
single-dish CO (2-1) and CO (3-2) spectra.

\newpage

\section{Discussion}

\subsection{A kinematical model of H$_2$ emission from RAFGL 2688}

We now construct a detailed, empirical model of the H$_2$
kinematics of RAFGL 2688. In formulating this model, we are
guided by our observations 
that (1) the polar lobes are characterized by velocity
gradients in which the fastest 
moving material is found closest to the star, and the slowest moving
material is found at the tips of the H$_2$ emission regions; 
and (2) the H$_2$ emission seen projected
along the equatorial plane contains a substantial velocity
gradient east to west and a second, smaller velocity gradient in the E lobe 
that runs north to south.

\subsubsection{Polar lobe H$_2$ kinematics}

The Gaussian shapes of the spatially integrated H$_2$ emission line profiles
of the polar lobes of RAFGL 2688 (Fig.\ \ref{fig:2688spectra}a) are similar
to the molecular emission line profiles detected in the radio regime from
species such as CO and HCN (e.g., Bieging \& Nguyen-Q-Rieu 1996).  The H$_2$
line profiles can be understood in the context of the detailed
morphology of H$_2$ emission seen in the NICMOS image (Fig.\
\ref{fig:2688NICMOS}). In this image the N and S lobes appear as
limb-brightened cavities. The emission peaks of the line profiles correspond
to the brightest regions of the lobes --- their limbs (i.e., outlines) ---
whereas the line wings are produced by emission from within the fainter lobe
interiors, which is dominated by material in the near sides (blue wings) and
far sides (red wings) of the lobes. 

The peak of the line profile therefore indicates the line of sight component
of the H$_2$ velocity at the lobe limb, and we can infer the deprojected N
and S lobe H$_2$ velocities as a function of position given an estimate of
$i$, the inclination of the polar axis of the system out of the plane of the
sky. Previous results (Yusef-Zadeh et al.\ 1984; Sahai et al.\ 1998a)
suggest $i\sim15^\circ$. The velocities we measure at the tips of the lobes
($\sim 5$ km s$^{-1}$; Fig.\ \ref{fig:2688poles}) then suggest that the gas
in this region is expanding away from the central star at a velocity $\sim
20$ km s$^{-1}$. We further infer that the H$_2$ velocities in the regions
of the lobes within $\sim1''$ of the inferred position of the central star
rise to $\sim 70$ km s$^{-1}$.  

For simplicity, 
we assume this behavior can be described by an inverse power law of the form
\begin{equation}
  v(\Delta\alpha,\Delta\delta) = v_0 (\frac{r_0}{r})^p \sin{i}
\end{equation}
for (RA, dec) offsets ($\Delta\alpha,\Delta\delta$) that lie within
$45^\circ$ of the projected polar axis of RAFGL 2688, where
$v_0$ is the velocity at the tips of the polar lobes,
$v(\Delta\alpha,\Delta\delta)$ is the observed radial
velocity at projected angular
radius $r = ((\Delta\alpha)^2 + (\Delta\delta)^2)^{1/2}$
from the central star, 
and $i$ is the inclination of the polar axis out of the plane of the sky. 
Based on the foregoing discussion of line profiles, we
make the further simplification that the polar H$_2$ emission is dominated
by emission from the lobe limbs. 

\subsubsection{Equatorial plane H$_2$ kinematics}

{\it If one postulates that all of the east-west H$_2$ emission is
confined to a narrow region along the equatorial plane,}
then a component of azimuthal velocity must be present in
the material responsible for this emission. Such a structure
might be analogous to that of the equatorial ``skirt'' observed in 
$\eta$ Car (Smith, Gehrz, \& Krautter 1998). Although the 
kinematics of the two objects appear to differ in detail
(Zethson et al.\ 1999), the H$_2$ morphology in 
the NICMOS image (Fig.\ 1) is consistent with the
confinement of the H$_2$-emitting material in RAFGL 2688 to
the plane of the system: knot E2 is brighter 
than knot E4, as expected if the former is directed toward the observer and
the latter is directed away from the observer; while E1 and E3,
which mark the perpendiculars to the polar axis (see
Weintraub et al.\ 2000, their Fig.\ 6), are of
similar intensity, as expected if we view these structures
through similar columns of intervening material.

We therefore assume that the
equatorial plane H$_2$ kinematics consist of a 
combination of azimuthal (rotation) and radial (expansion) velocity
components. We calculate these velocities from the
relation 
\begin{equation}
  v(\Delta\alpha,\Delta\delta) 
  = (v_e \sin{\phi} + v_r \cos{\phi}) \cos{i}
\end{equation}
for (RA, dec) offsets ($\Delta\alpha,\Delta\delta$) that lie within
$45^\circ$ of the projected equatorial plane of RAFGL 2688, where
$v_e$ is the expansion velocity, $v_r$ is the rotational velocity about the
polar axis, and $\phi$ is the azimuthal angle with respect to the polar axis
of RAFGL 2688, given by
\begin{equation}
  \phi = \tan^{-1} {(\frac{\Delta x}{\Delta y \; \sin{i}})}.
\end{equation}
Offsets ($\Delta x, \Delta y)$ are then related to
($\Delta\alpha,\Delta\delta$) via a rotation of the coordinate system by the
projected position angle of the polar axis on the plane of
the sky. 

\subsubsection{Comparison with data}

Model parameters $v_0$ and $p$ are relatively well constrained by Fig.\
\ref{fig:2688poles}, from which we infer $v_0 \sim 20$ km s$^{-1}$ and $p
\sim 0.7$. It is more difficult to constrain the values of the parameters
governing the equatorial plane H$_2$ radial velocities, $v_e$ and $v_r$. To
do so, we compared model velocity field images
($v(\Delta\alpha,\Delta\delta)$) calculated for a range of values of $v_e$
and $v_r$ with velocity centroids calculated from the Phoenix data cube
(see, e.g., Bieging \& Rieu 1996). We find qualitative and rough
quantitative agreement between model and data for values of
both $v_e$ and $v_r$
in the range $5-10$ km s$^{-1}$, with the additional constraint $v_e
+ v_r \sim 15$ km s$^{-1}$. An example of the results for a representative
successful model is displayed in Fig.\ \ref{fig:2688vcomp}. In this figure
the model velocity centroid image is calculated from Eqs.\ 1 and 2 for
values of $v_e = 5$ km s$^{-1}$ and $v_r = 10$ km
s$^{-1}$. 
There is clear qualitative agreement 
between the model and observed velocity images for these parameter values,
in the sense that the overall distribution of redshifted and blueshifted
emission is captured by the model. Furthermore, this model reproduces the
important details of the observed H$_2$ velocity distribution previously
described. Specifically, the model recovers the magnitudes and positions of
the observed blueshifted velocity extrema near (RA,dec) offsets
$(+1'',+2'')$ in the N lobe and $(+5'',-3'')$ in the E lobe, as well as the
magnitudes and positions of redshifted velocity extrema near offsets
$(-1'',-2'')$ in the S lobe and at $(-4'',+3'')$ in the W lobe.

Empirically, we find these parameter values --- $v_0=20$ km s$^{-1}$,
$p=0.7$, $v_e = 5-10$ km s$^{-1}$, and $v_r = 5-10$ km s$^{-1}$ --- provide
the best agreement between model and data in terms of overall qualitative
appearance of the velocity images as well as in the location {\it and}
magnitude of the velocity extrema. However, this model 
cannot be considered a ``best fit'' to the data. Rather, the foregoing
comparison of calculated and observed velocity fields
offers an indication of the magnitude of an azimuthal velocity
component relative to the components of radial expansion both parallel and
perpendicular to the polar axis.

\subsection{Alternative models}

The kinematic model described in \S 4.1
is not unique. For example, on the basis of
high spatial and spectral resolution CO
maps, which display kinematics very similar to those seen in H$_2$
emission, Cox et al.\ (1997, 
2000) and Lucas et al.\ (2000) have proposed that the
velocity gradients parallel and perpendicular to the
polar axis result from a multipolar system of jets whose directions of
motion are everywhere directed {\it radially outward} from the central
star. Since the entire fan-like eastern H$_2$ emission region
is blueshifted, the Cox et al.\ model requires that these
outflows must be directed well ``above'' the 
equatorial plane of the system, toward the observer, while the outflows
responsible for the western H$_2$ emission region are more tightly confined
and directed ``below'' the equatorial plane, away from the
observer.  Hence, in this model, the remarkable
orthogonality of the H$_2$ and CO emission morphologies is
purely a result of viewing angle. Multiple symmetry axes are
also suggested by observations of radio continuum emission
(Jura et al.\ 2000). Chaotic precession of bipolar jets might explain these
observations (see, e.g., Livio 2000 and references
therein). The very similar linear extents of the
orthogonal N-S and E-W molecular emission regions 
appear to pose a problem for any model invoking precessing
jets, however. 

Bieging \& Nguyen-Q-Rieu (1988, 1996) have proposed a model
in which the radio molecular line emission from the projected
equatorial plane of RAFGL 2688 originates in an
expanding, rotating equatorial disk or torus. Though qualitatively
similar to the model proposed in \S 4.1, our model
does not require that the H$_2$ emission arises from a disk.
Rather, the E-W H$_2$ lobes, like the polar (N-S) H$_2$ lobes (see
below), trace the interaction between
material ejected along the equatorial plane at two different
epochs; indeed, the 
interferometric radio molecular line observations  
indicate that the molecular jets detected in CO may be
responsible for shocks that produce the H$_2$
emission, based on the striking correspondence between
the CO and H$_2$ emission morphologies (Lucas et al.\
2000). There is no particular reason to assume that the
previously ejected material forms a complete torus around
the star and, indeed, Cox et al.\ (2000) and Morris \& Sahai
(2000) have argued, on the
basis of radio and mid-infrared imaging, respectively, that no such
structure exists. However, the optical and
infrared morphology of the nebula --- in particular, the
apparent occultation of the southern 
H$_2$ lobe and reflection nebulosity 
within $\sim2''$ of the central star --- does suggest the
presence of a large ($\sim5''$ radius), optically thick
structure. 

\subsection{Origin of the shocked H$_2$ emission}

\subsubsection{The transition from spherical to axial symmetry}

As noted by Sahai et al.\ (1998a), the detailed morphology of near-IR H$_2$
emission from RAFGL 2688 strongly supports previous suggestions that the
emission arises in shocks formed by the interaction of recently-developed,
fast-moving winds with slower-moving material. This relatively slow-moving
material likely was ejected during the AGB phase of the progenitor star.
Assuming J shocks are the origin of the gas heating that leads to H$_2$
emission then, according to model calculations (e.g., Burton, Hollenbach, \&
Tielens 1992), the shock fronts at the H$_2$ lobe tips are moving at a
minimum velocity $v_s \sim 10$ km s$^{-1}$ in the frame of the preshocked
(AGB) wind. This limit corresponds roughly to the energy threshold for
excitation of the $v=1-0$ $S(1)$ H$_2$ line. Hence, if the velocities
inferred for the tips of the polar lobes and for the
equatorial region, $\sim 20$ km s$^{-1}$ and $\sim 15$ km
s$^{-1}$, respectively, consist
of a superposition of (minimum) shock and AGB wind velocities, we infer that
the AGB wind was ejected at $v_{\rm AGB} \sim 5-10$ km s$^{-1}$, which is
typical of {\it low-mass} carbon stars. On the other hand, the nascent bipolar
structure and strong H$_2$ emission of RAFGL 2688 hint at a relatively
massive progenitor (Kastner et al.\ 1996) which may, in turn, suggest a
larger AGB outflow velocity ($v_{\rm AGB} \sim 20$ km s$^{-1}$; e.g.,
Kastner et al.\ 1993). If so, it would appear that, at their leading edges,
the shocks which excite H$_2$ have imparted very little momentum to the
former AGB wind.  In either case it is apparent that, well behind the
leading edges of the shocks that excite the H$_2$ in the
polar lobes, considerable wind
acceleration has taken place. The maximum speed for non-dissociative shocks
is $\sim25$ km s$^{-1}$ (Burton et al.\ 1992). Thus, within $\sim1''$
($\sim1000$ AU) of the central star of RAFGL 2688, the outflow velocity of
the gas is at least $\sim45$ km s$^{-1}$, which is much larger
than the outflow velocities of AGB stars. In RAFGL 618, meanwhile, the
inferred polar outflow velocities very near the central star are at least
$\sim100$ km s$^{-1}$ (neglecting projection effects), assuming the shocks
are traveling at $\sim25$ km s$^{-1}$.

We conclude that the velocity gradients along the polar axes of both RAFGL
2688 and RAFGL 618 trace rapid transitions from the ``slow,''
spherically symmetric winds of their AGB progenitors to faster, collimated,
post-AGB winds. A likely mechanism for such a transition is that late AGB
and post-AGB ejecta relatively close to the central star are accelerated
with increasing efficiency as the star is ``unveiled'' and circumstellar
material is exposed to a progressively hotter photosphere and, potentially,
a stronger stellar magnetic field.  Indeed, RAFGL
2688 has been observed to be increasing steadily in optical magnitude
(Gottlieb \& Liller 1976). These conditions would appear to be conducive to
the formation of H$_2$ shocks during the post-AGB but pre-planetary nebula
phase of bipolar nebulae, a scenario consistent with the conclusions of
Weintraub et al.\ (1998). 
Such shocks could then play a central role in the
transition from spherically symmetric to axisymmetric outflow, as the
``young,'' fast (but still predominantly molecular) wind driven by the
emerging post-AGB star excavates bipolar cavities within the previously
ejected red giant wind. 
The displacement from the central
star to the tips of the H$_2$ lobes ($\sim6000$ AU), combined with the
deprojected velocities at the lobe tips, $v_0 \sim 20$ km s$^{-1}$,
suggests the H$_2$ lobes have a dynamical age of $1500$ yr, which then
serves as an upper limit to the post-AGB ``unveiling'' of the central star.

\subsubsection{The Egg Nebula as endpoint of common envelope binary evolution}

A presently popular theory accounts for the structure 
of bipolar planetary nebulae as due to the interaction of a close binary
system, in which the secondary diverts mass loss from and/or spins up the
envelope of the primary (e.g., Soker 1997, 1998).  
According to a specific formulation of the binary
progenitor model (Soker 1998), the relatively massive progenitors of bipolar
PNs undergo a phase during the AGB evolution of the
primary in which the secondary resides just outside the envelope of the
primary and diverts mass loss into the equatorial plane. This phase is
followed by a short-lived common envelope phase that, presumably, rapidly
spins up the envelope of the primary; the carbon star V Hya, which displays
$V \sin{i} \sim 13$ km s$^{-1}$, may be undergoing such a phase at present
(Barnbaum et al.\ 1995). 

Many, though certainly not all, of the observed characteristics of
RAFGL 2688 appear to be explained by such a
model. Specifically, its notorious system of 
concentric broken arcs would result from episodic, spherically
symmetric mass loss during a prolonged period of
quasi-single star AGB mass loss 
preceding the common envelope phase; the subsequent
formation of bipolar structure would be the result of the
increasingly close proximity of a companion during the late AGB
evolution of the primary; and the H$_2$ shocks would mark the 
rapid termination of a common envelope phase and the resultant 
sudden ejection of the remaining AGB star envelope. 
Hence, the H$_2$ emission regions are produced by bullet-like
ejecta that have collided with material 
that was recently ejected --- much of it along the equatorial plane 
--- during the close binary phase preceding common envelope formation. 

It would appear that no existing model can explain
satisfactorily the rich phenomenology of the Egg Nebula. The
molecular kinematics of its equatorial region remain
particularly problematic for present models, whether or not
a component of azimuthal velocity is present along the
equatorial plane of the system. A pure radial outflow model
cannot explain the striking orthogonality of the molecular
emission from RAFGL 2688; furthermore, such a model is, at
present, reliant on {\it ad hoc} physical mechanisms for the
presence of multiple symmetry axes (e.g., Frank 2000). On
the other hand, the azimuthal velocities of the equatorial
H$_2$ emission suggested by our model and the models of
Bieging \& Nguyen-Q-Rieu (1988, 1996) are far too large to
be explained by Keplerian rotation; an alternative source of
angular momentum is required.  As proposed by Bieging \&
Nguyen-Q-Rieu, this angular momentum might be provided by
magnetic fields that are corotating with the star. It is of
considerable importance to establish whether sufficiently
large fields can be generated via interaction with a close
companion star (Bieging \& Nguyen-Q-Rieu 1988, 1996) or by
dynamo activity in the primary post-AGB star itself (e.g.,
Soker 2000; Blackman, Frank, \& Welch 2000). 
The latter mechanism could be particularly
effective if the central star has just undergone a common
envelope phase, such that highly magnetically active
interior layers of the star drive the molecular jets seen in
CO and H$_2$.

\acknowledgements{Support for this research was provided in
part by a JPL/ISO grant to R.I.T. We acknowledge enlightening discussions
with Adam Frank and Robert Lucas, and the helpful comments
of the referee.}

\newpage

\section*{Figure captions}

\begin{enumerate}

\item 
\label{fig:2688NICMOS}
HST/NICMOS image of 2.12 $\mu$m H$_2$ emission from RAFGL
2688 (Sahai et al.\ 1998a). Features identified E1--E4 by Sahai et al.\ are
indicated.

\item 
\label{fig:2688posvel}
Phoenix spectral images of RAFGL 2688. Images were obtained at the
declination offsets indicated above each frame. Each calibrated image is
represented as a position-velocity map, with RA running along image columns
and radial velocity running along image rows. The velocity scale of each
image is centered on the systemic velocity of RAFGL 2688 with respect to the
local standard of rest, $V_{sys} = -34$ km s$^{-1}$ (e.g., Bieging \&
Nguyen-Q-Rieu 1996). East is to the left; the spatial scale is
$0.35''$ per pixel, with the central star located very
near pixel column 42 (nebular continuum emission is also evident
in the images in the center panels). Relative  
intensities in these and other images presented in this
paper are represented with a linear greyscale 
unless otherwise indicated.

\item 
\label{fig:2688spectra}
Spectra extracted {\it a)} from spectral images obtained at declination
offsets of $+4''$ (N lobe) and $-4''$ (S lobe) and {\it b)} from spectral
images obtained at $0''$ declination offset (E and W lobes). Each spectrum
is obtained by summing image sections of width $4.5''$ in RA centered on the
peak of H$_2$ emission. The velocity scale of each spectrum is centered on
the systemic velocity of RAFGL 2688 with respect to the local standard of
rest. 

\item 
\label{fig:2688cube}
Velocity-integrated H$_2$ images of RAFGL 2688, extracted from the Phoenix
data cube. Row and column image scales indicate offsets
(arcsec) in RA and dec, respectively. Each image is
integrated over $\sim5$ km s$^{-1}$ 
and is centered at the velocity indicated. These velocities
are relative to the systemic velocity of RAFGL 2688.

\item 
\label{fig:2688poles}
Radial velocity centroid, with respect to systemic velocity, of H$_2$
emission vs.\ position along the polar axis of RAFGL 2688. Uncertainties in
determinations of velocity centroids are approximately the symbol size.

\item 
\label{fig:618posvel}
Calibrated Phoenix spectral image of RAFGL 618. The velocity scale of the
image is centered on the systemic velocity of RAFGL
618. East is to the left; the spatial scale is
$0.35''$ per pixel, with the central star located very
near pixel column 18 (continuum emission is present
at this position in the form of vertical band across the
image). The image is displayed in a 
logarithmic greyscale to bring out the faint line wing
emission, which extends to $\sim \pm 120$ km s$^{-1}$. 

\item 
\label{fig:618spectra}
Spectra extracted from the spectral image of RAFGL 618. Each spectrum
is obtained by summing image sections of width $2''$ in RA at the offset
indicated.  The velocity scale of each spectrum is centered on
the systemic velocity of RAFGL 618 with respect to the local standard of
rest ($-21$ km s$^{-1}$; e.g., Meixner et al.\ 1998). 

\item 
\label{fig:2688vcomp}
Comparison of observed and model H$_2$ velocity fields
for RAFGL 2688. The observed velocity field (left) consists
of velocity centroids calculated from the Phoenix data cube. In
the model (right), we set the equatorial expansion velocity at $v_e=5$ km
s$^{-1}$ and the equatorial rotation velocity at $v_r=10$ km s$^{-1}$ (see
\S 4.3). We display only those portions of the
model velocity image for which the SNR in the HST/NICMOS H$_2$ image exceeds
$\sim5$, and only those portions of the observed velocity centroid image for
which the SNR in the velocity-integrated Phoenix H$_2$ image exceeds
$\sim3$. 

\end{enumerate}

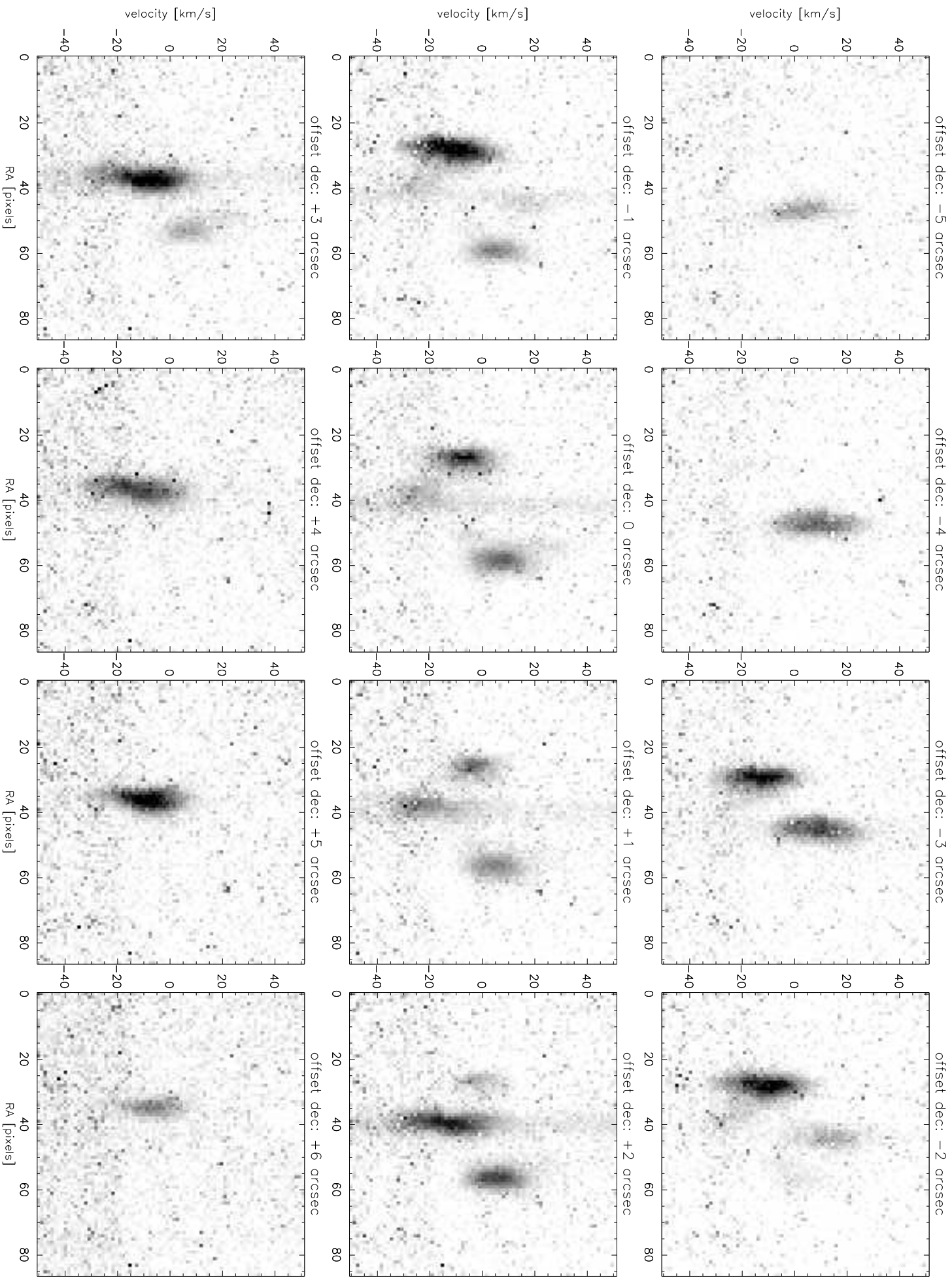
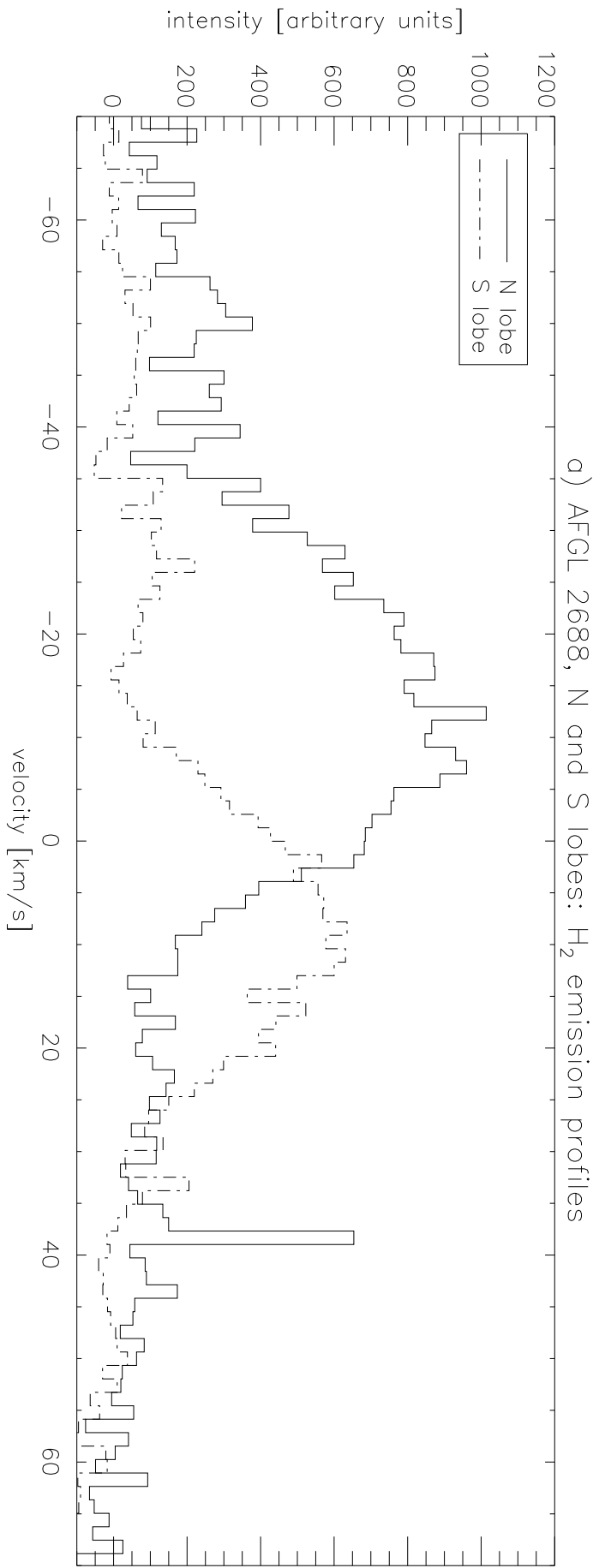
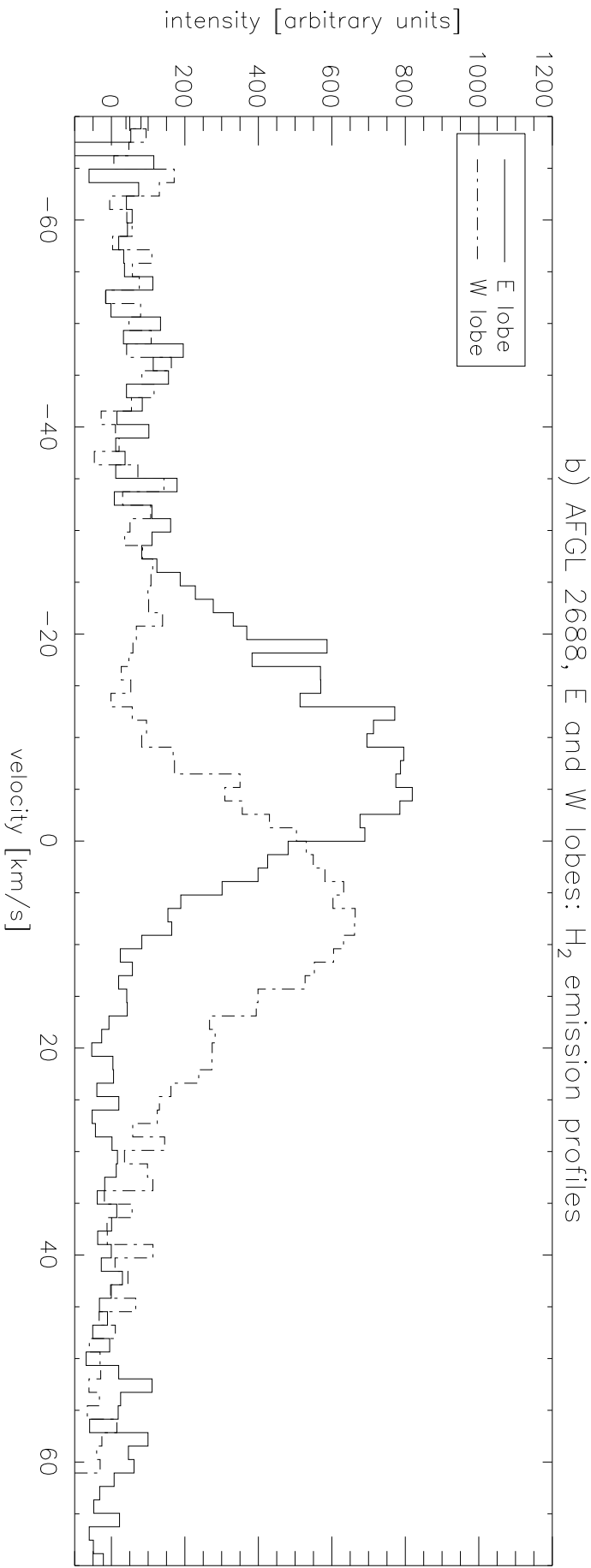
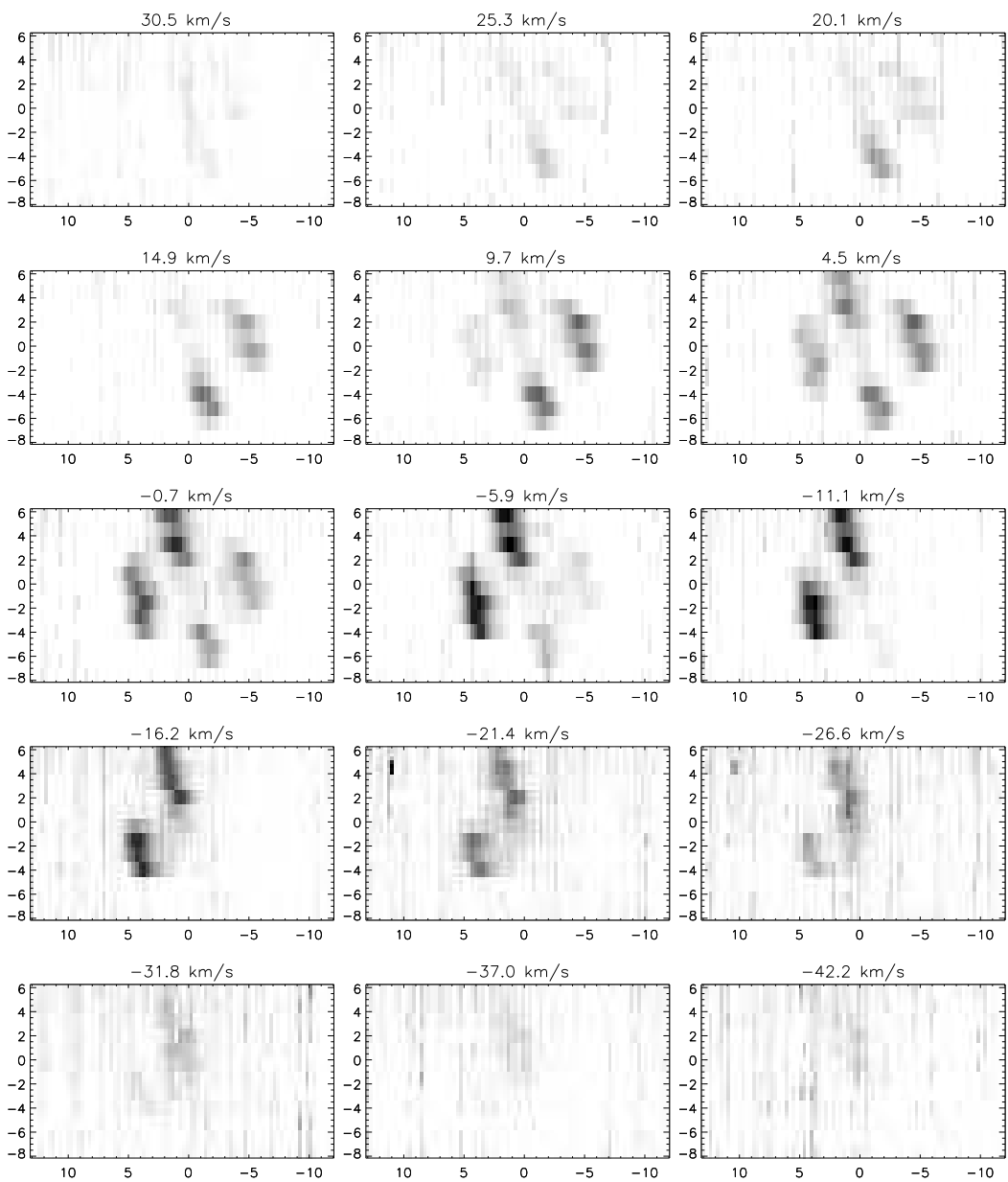
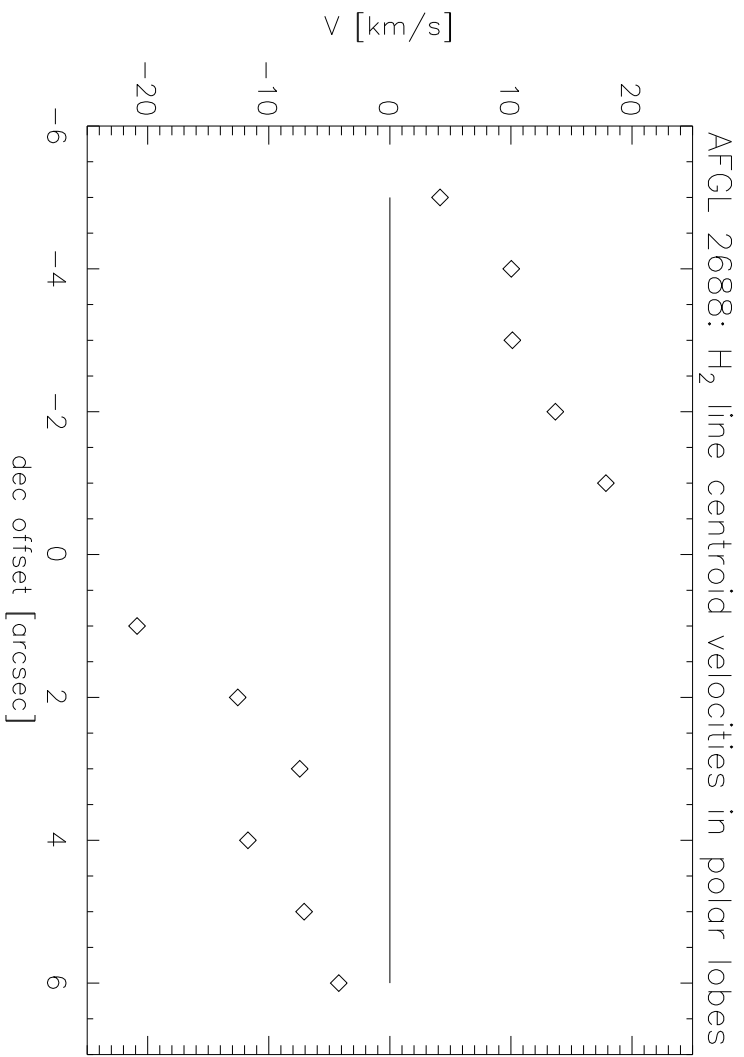
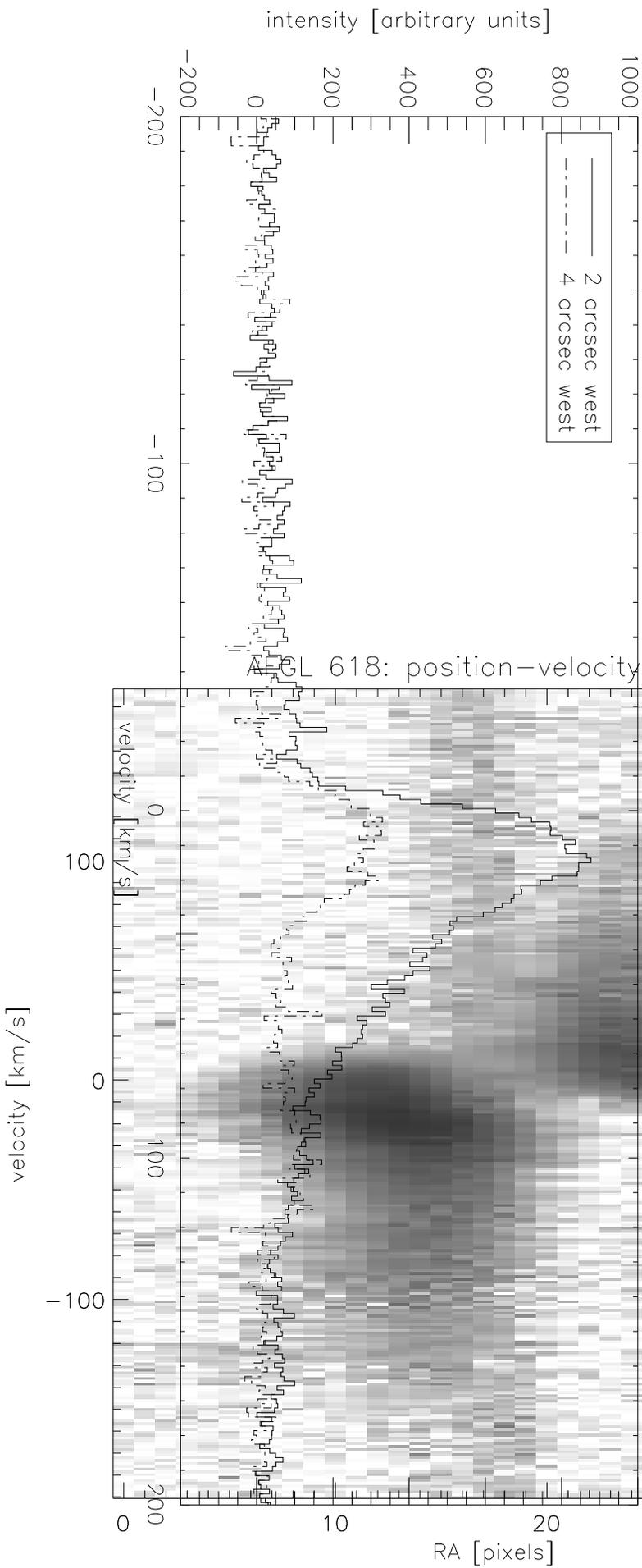
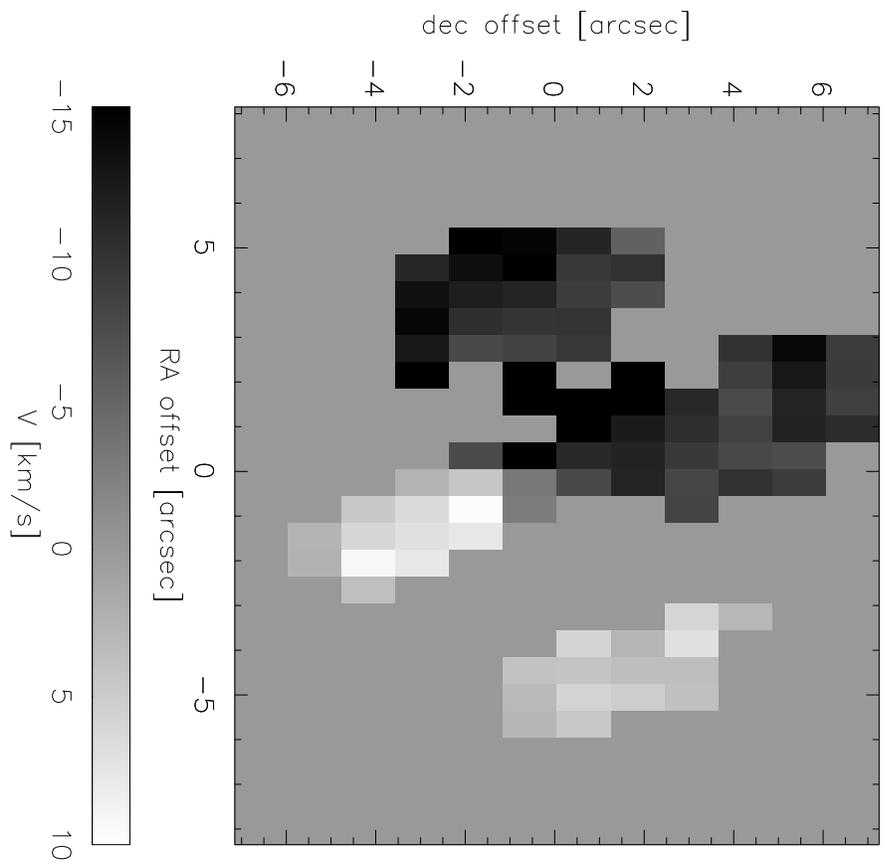
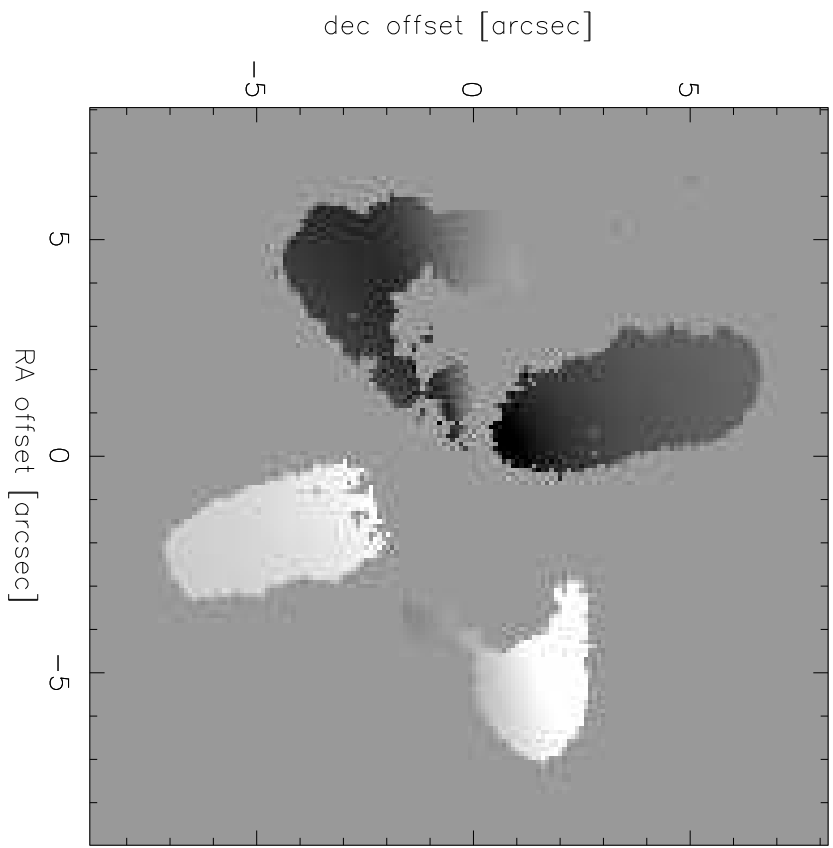
\end{document}